# LASER-ASSISTED DEPOSITION OF r-$B_4C$ COATINGS USING ETHYLENE AS CARBON PRECURSOR

M.J. Santos[1], A.J. Silvestre[2] and O. Conde[1*]

[1]*Departamento de Física, Universidade de Lisboa, Ed. C8, Campo Grande, 1749-016 Lisboa, Portugal.*

[2]*Instituto Superior de Transportes, R. Castilho, No.3, 1269-074 Lisboa, Portugal.*

## Abstract

Rhombohedral $B_4C$ coatings were synthesised on fused silica substrates by $CO_2$ laser-assisted chemical vapour deposition (LCVD), using a dynamic reactive atmosphere of $BCl_3$, $C_2H_4$ and $H_2$. Films with carbon content from 15 to 22 at.% were grown at deposition rates as high as 0.12 $\mu m.s^{-1}$. The kinetics of the reactive system used to deposit the $B_4C$ films and the influence of growth conditions on the structure and morphology of the deposits were investigated.

## 1. Introduction

Rhombohedral boron carbide (r-$B_4C$) is a ceramic material of great interest for a wide variety of applications because of its attractive mechanical, thermal and electronic properties [1]. The r-$B_4C$ has found application as a neutron absorbent material in nuclear industry due to its high neutron capture cross-section [2]. Also of particular importance are the low specific weight and high hardness, the latter even surpassing diamond and boron nitride at temperatures over 1100 ºC [3]. Moreover, it presents a high melting point and modulus of elasticity, and has great resistance to chemical agents. This combination of properties makes boron carbide a prominent corrosion-resistant ceramic material for thin film applications. Furthermore, considering its high-temperature stability, large Seebeck coefficient and low thermal conductivity, boron carbide could find potential use as high-temperature thermoelectric material for energy converters [1,4].

[*] Corresponding author: Tel. + 351 21 7500035, Fax: +351 21 7573619, E-mail: oconde@fc.ul.pt

In previous studies conducted by our group, laser-assisted chemical vapour deposition (LCVD) of boron-carbon films was carried out from a dynamic reactive atmosphere of $BCl_3$, $CH_4$, and $H_2$ using a continuous wave $CO_2$ laser [5-7]. Facing to methane, which is the conventional carbon precursor in CVD processes, ethylene presents several advantages due to its high absorption coefficient at the $CO_2$ laser wavelength and its higher sticking coefficient, enabling to achieve higher deposition rates.

It is the purpose of this paper to report on the kinetics of $CO_2$ laser-CVD of r-$B_4C$ films using $C_2H_4$ as carbon precursor and on the influence of growth conditions on the structure and morphology of the deposited material.

## 2. Experimental procedure

The deposition system and the procedure used for laser-assisted CVD of r-$B_4C$ have been presented elsewhere [5] and only a brief description is given here. Boron carbide films were deposited on silica substrates using a $CO_2$ laser as heat source, operated in cw $TEM_{00}$ mode at a wavelength of 10.6 μm, and a dynamic reactive gas mixture of $BCl_3$, $C_2H_4$ and $H_2$. Argon was used as buffer gas. The laser beam reaches the substrate at perpendicular incidence with a $1/e^2$ spot diameter of 12 mm. No focus lens was used since silica absorbs 84 % of the laser radiation. Prior to their insertion in the reactor, the substrates were cleaned in ultrasonic baths of acetone and ethanol. The reaction chamber was always evacuated to a base pressure of $<10^{-6}$ mbar before the introduction of the gaseous reactants.

In this study, the total pressure and argon flux were kept constant at 133 mbar and 400 sccm, respectively. The other experimental parameters were varied in the ranges shown in table 1. The relative amount of carbon and boron in the reactive gas phase is given by the parameter $\varphi = 2\Phi_{C2H4}/(2\Phi_{C2H4} + \Phi_{BCl3})$, where $\Phi_i$ is the flow rate of precursor i. In this study φ took values between 0.05 and 0.13.

**Table 1**

Process parameters for LCVD of r-$B_4C$ films

| *Experimental parameters* | *Range of values* |
|---|---|
| Laser power (W) | 180 - 250 |
| Interaction time (s) | 90 |
| $BCl_3$ flow rate (sccm) | 41 |
| $C_2H_4$ flow rate (sccm) | 1 - 3 |
| $H_2$ flow rate (sccm) | 150 - 200 |

The structure of the as-deposited films was studied by X-ray diffraction at glancing incidence (GIXRD) of 1º with Cu-Kα radiation and the chemical composition was investigated by electron probe microanalysis (EPMA). The surface microstructure of the films was examined by scanning electron microscopy (SEM) and thickness profiles were measured by optical profilometry.

## 3. Results and discussion

### *3.1 Chemical and structural analysis*

Chemical analysis by EPMA showed that films with uniform composition were produced in a broad range of carbon content, from 15 to about 22 at.% C, consistent with the B-C phase diagram.

All X-ray spectra display the r-$B_4C$ diffraction pattern, showing narrow diffraction lines.

Depending on the carbon content, the spectra match the JCPDS cards 33-0225 or 35-0798 corresponding to the $B_{13}C_2$ and $B_{12}C_3$ stoichiometries, respectively (Fig. 1). Moreover, the deposition of graphite was never observed and only boron carbide phase was detected by GIXRD.

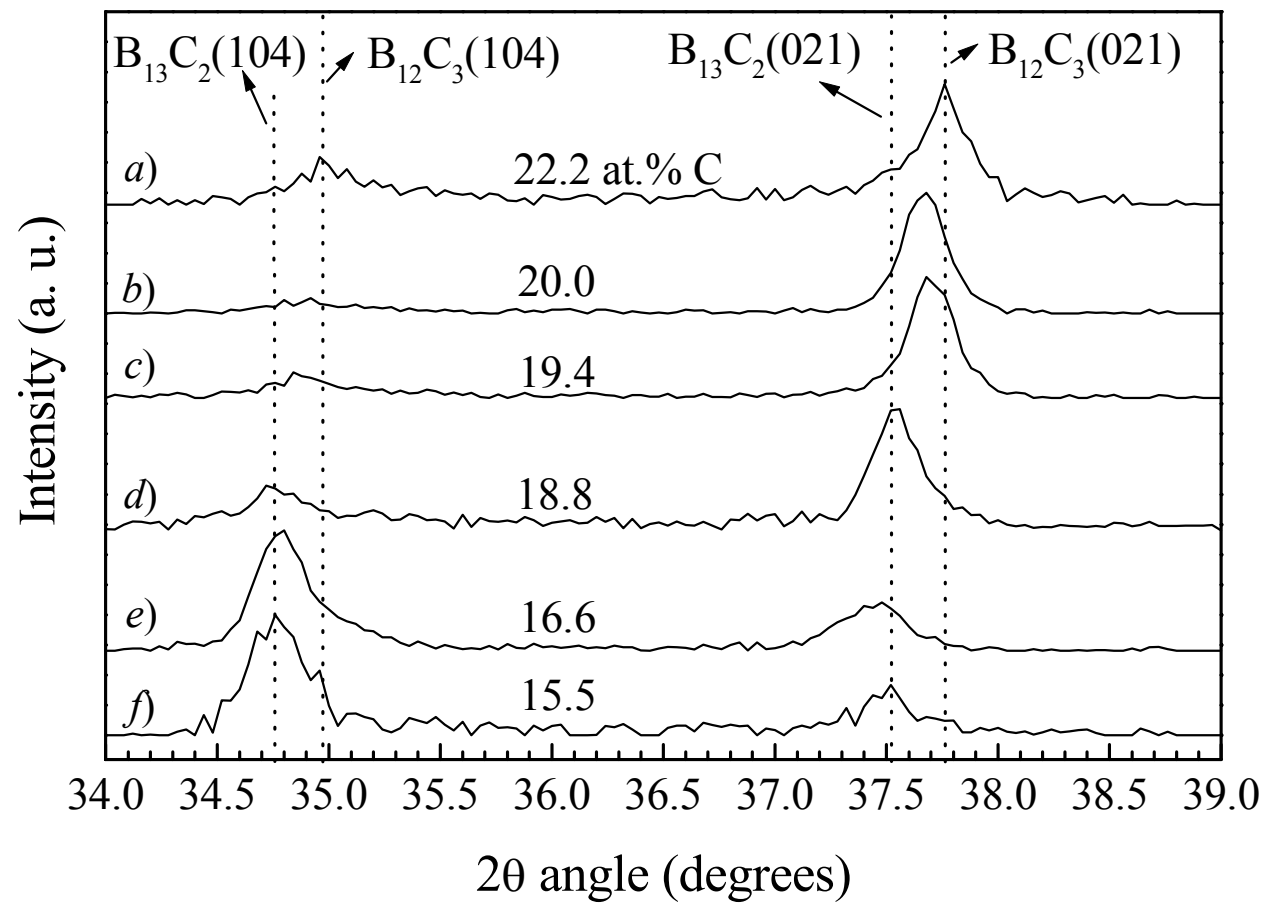

Fig. 1. GIXRD spectra of r-$B_4C$ films with different carbon contents, as measured by EPMA, prepared with $C_2H_4$ at the following experimental parameters: a) $\varphi$=0.13, $\Phi_{H2}$=200 sccm, P=220 W; b) $\varphi$=0.13, $\Phi_{H2}$=150 sccm, P=220 W; c) $\varphi$=0.09, $\Phi_{H2}$=200 sccm, P=240 W; d) $\varphi$=0.09, $\Phi_{H2}$=150 sccm, P=200 W; e) $\varphi$=0.05, $\Phi_{H2}$=150 sccm, P=240 W; f) $\varphi$=0.09, $\Phi_{H2}$=200 sccm, P=200 W.

The two major diffraction peaks characteristic of r-$B_4C$ compound correspond to the (104) and (021) reflections. Their 2θ angular position varies with the carbon content in the films, as can be seen from Fig. 1. Both peaks shift left approximately 0.2° when the carbon concentration decreases from about 22 to about 15 at.% C. This observed shift is mainly related with the substitution of a carbon atom by a boron atom in the central C-B-C intericosahedral chain ($B_{12}C_3$ structure), leading to a C-B-B chain in the main diagonal of the rhombohedral structure of $B_{13}C_2$ boron carbide [7].

Similarly to the films synthesised with methane [6,7], the coatings prepared from ethylene present an inversion of the relative intensities of the (104) and (021) lines for carbon content lower than 17 at.% C, suggesting the development of a (104) texture (Fig. 1, diffractograms e and f). Although this trend to develop the (104) texture is not well understood, it is clearly independent of carbon precursor. Furthermore, preferential crystallographic orientations in this range of carbon content have also been observed in other CVD processes, namely in low-pressure CVD [8].

*3.2 Growth kinetics*

Gaussian or near flat thickness distribution profiles were observed, allowing to calculate apparent deposition rates by taking the ratio of the maximum height, above the substrate surface, to the irradiation time. Deposition rates between 0.03 and 0.12 $\mu m.s^{-1}$ were measured, which are one order of magnitude larger than those obtained in CVD processes [9,10]. Fig. 2 shows the deposition rate of r-$B_4C$ films as a function of $H_2$ flow rate, for $\varphi$=0.09 and $\varphi$=0.13 and for three distinct laser power values. As can be seen, the deposition rate slightly increases with the $\varphi$ parameter while laser power and hydrogen concentration in the gaseous mixture are the process parameters that most strongly determine the rate at which films are grown. For a given reactive atmosphere/substrate system and interaction time, the maximum temperature attained during the laser-material interaction increases as laser power increases. As a consequence, the deposition rate is greatly influenced by this process parameter.

As also depicted in both graphs in fig. 2, the deposition rate varies with $\Phi_{H2}$ going through a maximum at 175 sccm. The increase of the deposition rate with the hydrogen flux, in the low $\Phi_{H2}$ range, is expected because hydrogen favours the deposition reaction of boron carbide by reducing the $BCl_3$ and preventing graphite formation. On the contrary, the decreasing behaviour for the higher $\Phi_{H2}$ values could be explained by considering that the deposited thin films are etched by the hydrogen. Nevertheless, the usual temperature for this process in bulk $B_4C$ is ~1200 ºC [11], which is much higher than the deposition temperature values calculated in this work. Another hypothesis to understand the decrease of the deposition rate at high $\Phi_{H2}$ is based on Soret's effect, by which the heavier $BCl_3$ and $C_2H_4$ molecules are relegated to the colder zones, a substantial fraction of the adsorption sites in the central irradiated region being occupied by the lighter $H_2$ molecules [12]. This leads to a limited access of the B and C precursors to the central reaction zone, and therefore to an inhibition of film growth. The balance between these two processes yields a $\Phi_{H2}$ value at which the deposition rate of r-$B_4C$ presents a maximum. For this optimised $H_2$ flux, the number of nucleation centres is higher leading not only to higher deposition rates but also to films presenting a much more compact and uniform microstructure.

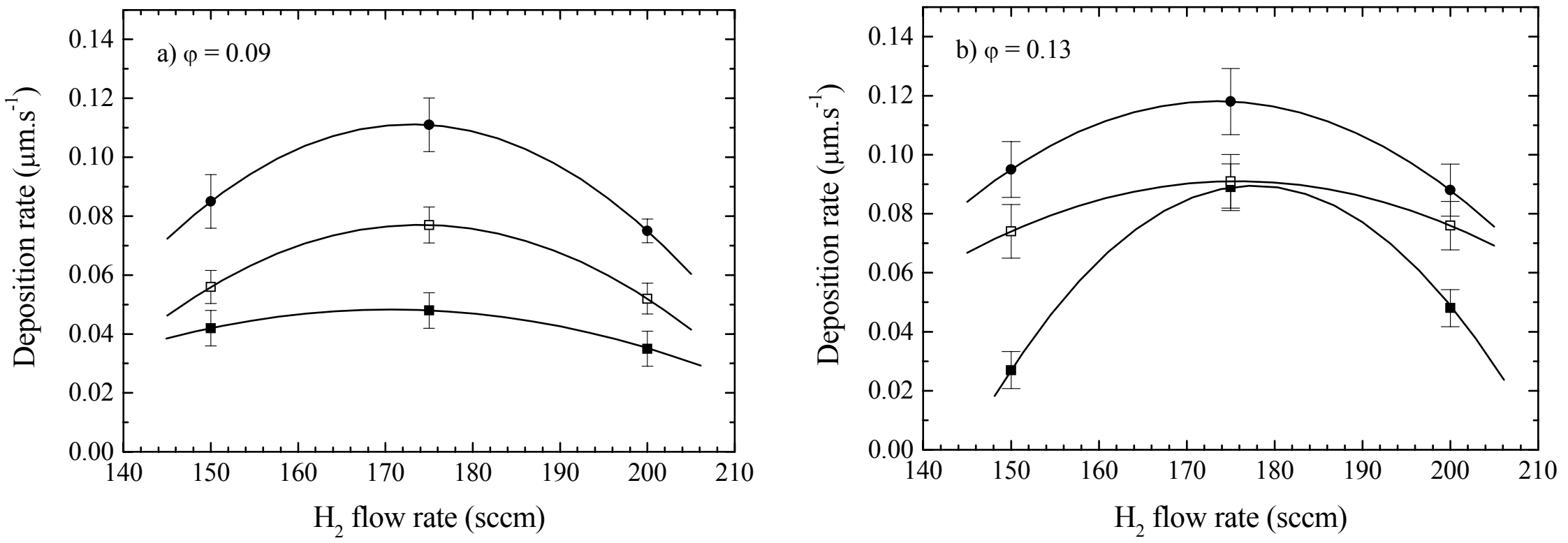

Fig. 2. Deposition rate as a function of $H_2$ flow rate, for two different φ values and for the following laser power values: ■, P=200 W; □, P=220 W; •, P=240 W.

Arrhenius diagrams are a common convenient way to infer on the reaction mechanism that controls deposition [13]. Mass transport in the vapour phase and surface kinetics are the rate-limiting mechanisms in conventional CVD and also in LCVD, although the latter process has a more complex structure. The Arrhenius plot presented in Fig. 3 shows the *ln* of the r-$B_4C$ deposition rate as a function of the reciprocal central deposition temperature. The surface temperature achieved at the centre of the films during deposition was estimated between 780 K and 1430 K, following the calculation technique described in ref. [6]. It can be seen from the Arrhenius diagram that two straight lines with different slopes are needed to fit the data. Thus, within the deposition temperature range induced by the laser radiation and for the reactive atmosphere utilised, two rate-limiting mechanisms were identified for the LCVD of r-$B_4C$ films:

*i)* At high deposition temperatures (T>1050 K), the rate-limiting step for film growth is the mass transport of the reactive gaseous species. In this temperature region the slope of the straight line has a small value of 5.3 $kJ.mol^{-1}$. Due to the reduced number of experimental data used in the fitting procedure, a high standard error of ± 2.7 $kJ.mol^{-1}$ was calculated.

*ii)* At low deposition temperatures (T<1050 K), the apparent activation energy deduced from the Arrhenius plot is $E_a$=30.7±2.1 $kJ.mol^{-1}$. We can thus conclude that surface chemical reaction kinetics is the rate-limiting step for film growth at temperatures below 1050 K.

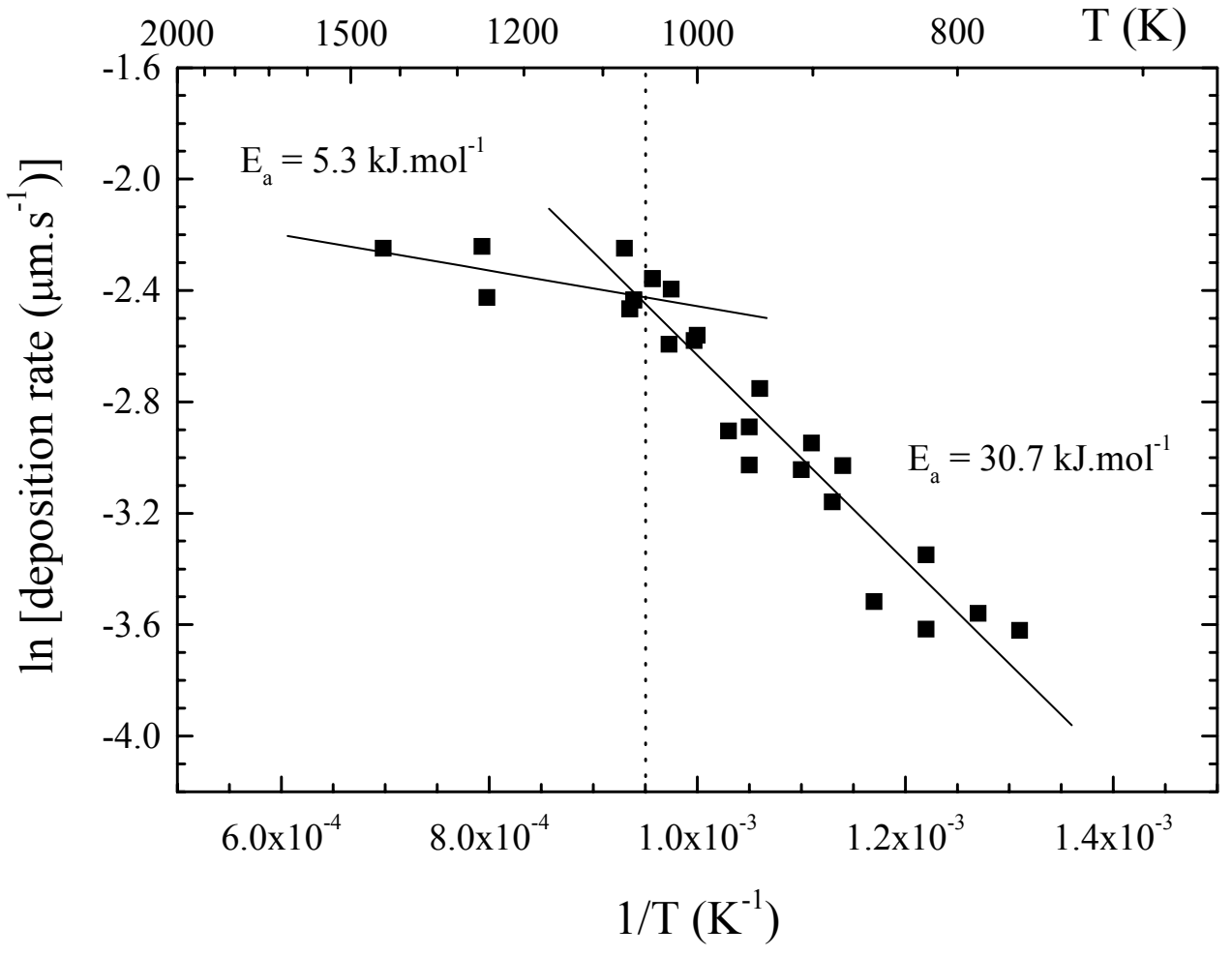

Fig. 3. Logarithm of the apparent deposition rate as a function of reciprocal temperature. ■, measured values; —, least square fitting.

It should be noted that the magnitude of the apparent activation energy deduced for the kinetic regime is much lower than the $E_a$ values that usually characterise this regime in purely thermal CVD processes [13]. Because ethylene, as well as boron trichloride, absorbs the infrared $CO_2$ laser radiation through vibrational molecular excitation [14-16], different reaction pathways can be opened between the excited reactants leading to a lower reaction activation energy.

*3.3 Microstructure*

SEM analysis of the coatings surface shows a crystalline morphology, the size and shape of the crystals depending on the deposition conditions. Fig. 4 illustrates the microstructure evolution with both laser power and gas phase composition. Also a cross-section of a coating is presented exhibiting good adherence and columnar growth. As can be seen in Fig. 4, sequence a), hydrogen concentration in the gaseous mixture plays an important role in the microstructure. Following the discussion given in the previous section, the more compact and uniform microstructure is achieved for the optimised $H_2$ flux, i.e. 175 sccm. The microstructure is also strongly determined by the relative amount of carbon and boron in the reactive atmosphere. Fig. 4b) shows the evolution as $\varphi$ increases from 0.05 to 0.13. At low $\varphi$ values, the growth of approximately spherical nodules is observed. For higher carbon content in the gas phase, a second nucleation and growth mechanism takes place leading to a fine and uniform grain structure that covers the underlying nodular structure. At maximum $\varphi$ values, nodules disappear and a continuous and uniform morphology develops showing well-defined crystal facets with a pyramidal geometry. Moreover, SEM micrographs reveal a substantial growth of the grain size (Fig. 4c) as laser power increases by only 20 W (from 220 to 240 W) emphasising the remarkable influence of the laser power/deposition temperature as referred to previously.

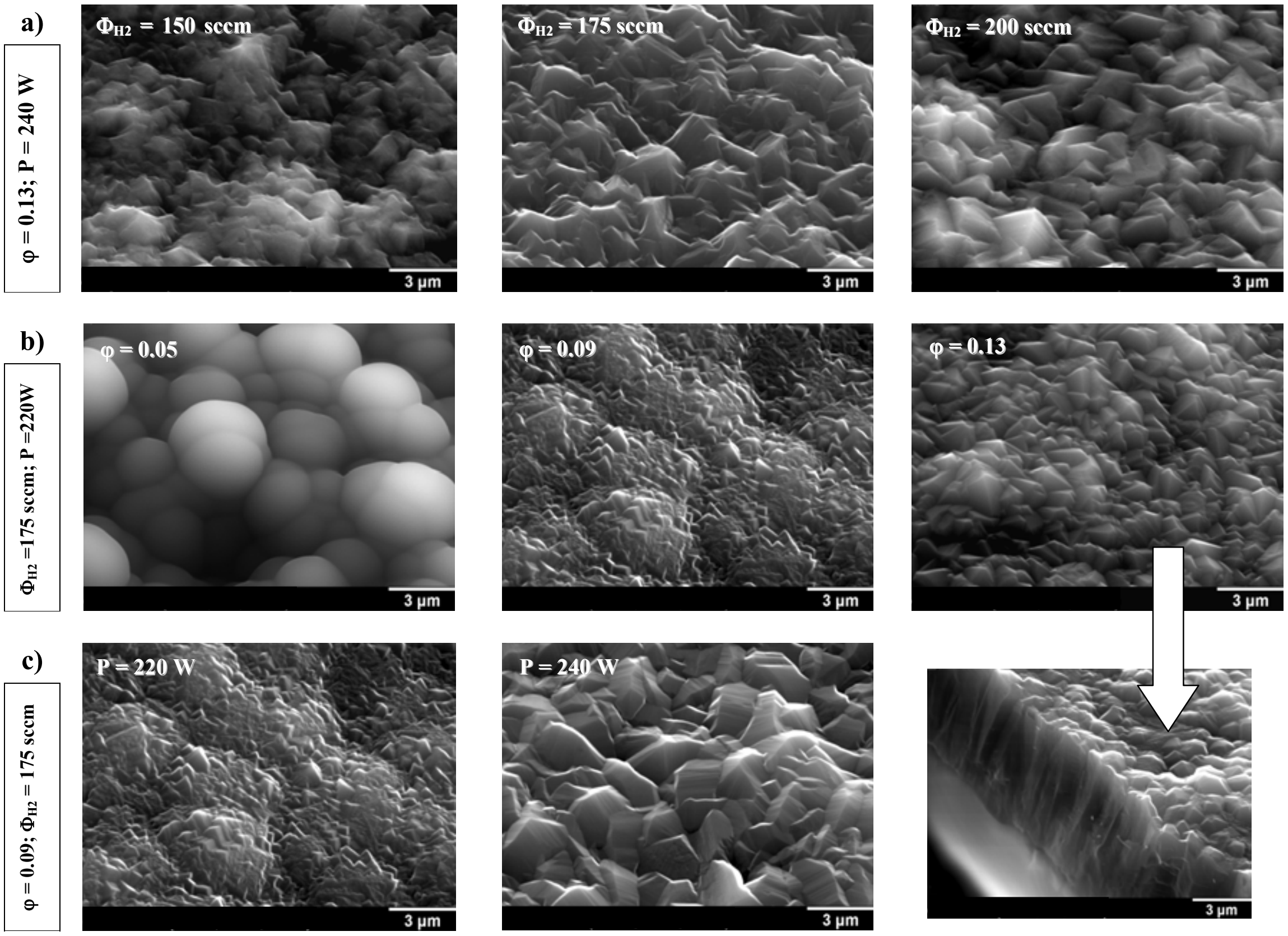

Fig. 4. Evolution of surface microstructure of films processed at different experimental conditions.

## 4. Conclusions

Laser-CVD from $BCl_3$ and $C_2H_4$ as boron and carbon precursors, respectively, yield r-$B_4C$ films with good adherence, well developed grain structure and carbon content in the range 15 to 22 at.%. The deposition rate and the surface microstructure strongly depend on laser power and hydrogen content in the gas phase. For the experimental conditions used in this work, it was shown that growth kinetics is dominated by a surface reaction mechanism at T<1050 K while mass transport in vapour phase is the controlling mechanism of thin film growth at T>1050 K.

## Acknowledgements

This work was partially funded by Fundação para a Ciência e Tecnologia under POCTI.

### References

[1] D. Emin, Phys. Rev. B 38 (1988) 6041.

[2] X. Deschanels, D. Simeone and J.P. Bonal, J. Nucl. Mater. 265 (1999) 321.

[3] R. Telle, in: M. V. Swain (Ed.), Structure and Properties of Ceramics, Mater. Sci. Technol., vol. 11, VCH Publishers, Weinheim, 1994, p. 173.

[4] M. Olsson, S. Soderberg, B. Stridh, U. Jansson, J.-O. Carlsson, Thin Sol. Films 172 (1989) 95.

[5] J.C. Oliveira, M.N. Oliveira, O. Conde, Surf. Coatings Technol. 80 (1996) 100.

[6] J.C. Oliveira, O. Conde, Thin Sol. Films 307 (1997) 29.

[7] O. Conde, A.J. Silvestre, J.C. Oliveira, Surf. Coatings Technol. 125 (2000) 141.

[8] J. Rey, G. Male, Ph. Kapsa, J.L. Loubet, J. Physique Coll. C5 50 (1989) 311.

[9] U. Jansson, J.-O. Carlsson, B. Stridh, J. Vac. Sci. Technol. A5 (1987) 2823.

[10] O. Postel, J. Heberlein, Surf. Coatings Technol. 108-109 (1998) 247.

[11] K.A. Schwetz and A. Lipp, in: Ullmann's Encyclopedia of Industrial Chemistry, vol. A4, VCH Verlag, Weinheim, 1985, pp. 295–307.

[12] J.O. Hirschfelder, C.F. Curtiss and R.B. Bird, Molecular Theory of Gases and Liquids, M.G Mayer (Ed.), John Wiley & Sons, New York, 1964.

[13] J.-O. Carlsson in: D.S. Rickerby and A. Matthews (Eds.), Advanced Surface Coatings – A Handbook of Surface Engineering, Blackie & Son, 1991, pp. 162–193.

[14] N. Karlov, Appl. Optics 13 (1974) 301.

[15] T. Shimanouchi, Tables of Molec. Vibrational Frequencies Consolidated, vol. I, NSRDS-NBS 39, 1991, p. 74.

[16] J.I. Steinfeld, Laser and Coherence Spectroscopy, Plenum Press, New York, 1978, p. 62.